\documentclass[useAMS,usenatbib,referee]{biom}
\def\bSig\mathbf{\Sigma}

\usepackage{graphicx}
\usepackage[textwidth=8em,textsize=small]{todonotes}
\usepackage{amsmath}
\usepackage{natbib}
\usepackage{amsmath}

\usepackage{times}
\usepackage{bm}
\usepackage{natbib}
\usepackage{amsmath}
\usepackage{graphicx}
\usepackage{enumerate}
\usepackage{hyperref}
\usepackage{xurl}
\usepackage{amssymb}
\usepackage{mathtools}
\usepackage{setspace}
\usepackage{array,booktabs}
\usepackage{tikz}
\usepackage{mathtools}
\usepackage{chngcntr}
\usepackage{algorithm}
\usepackage{algpseudocode}
\usepackage{multirow}
\usepackage{multicol}
\usepackage{subcaption}
\usepackage{amsmath}
\usepackage{footnote}
\usepackage{enumitem}

\DeclareMathOperator*{\argmin}{arg\,min}

\renewcommand{\thesection}{\arabic{section}}
\graphicspath{{./art/}}

\title[A general, flexible and harmonious framework for interpretation]{A general, flexible and harmonious framework to construct interpretable functions in regression analysis}

\author{Tianyu Zhan$^{1,*}$\email{tianyu.zhan.stats@gmail.com}\thanks{Co-corresponding author} and Jian Kang$^{2, **}$\email{jiankang@umich.edu}\thanks{Co-corresponding author} \\
	$^{1}$Data and Statistical Sciences, AbbVie Inc., 1 Waukegan Road, North Chicago, IL 60064, U.S.A.\\
	$^2$ Department of Biostatistics, University of Michigan, 1415 Washington Heights, Ann Arbor, MI 48109, U.S.A.}

\begin{document}





\pagerange{\pageref{firstpage}--\pageref{lastpage}} 
\volume{XX}
\pubyear{2021}
\artmonth{July}


\doi{10.1111/j.1541-0420.2005.00454.x}


\label{firstpage}


\begin{abstract}

An interpretable model or method has several appealing features, such as reliability to adversarial examples, transparency of decision-making, and communication facilitator. However, interpretability is a subjective concept, and even its definition can be diverse. The same model may be deemed as interpretable by a study team, but regarded as a black-box algorithm by another squad. Simplicity, accuracy and generalizability are some additional important aspects of evaluating interpretability. In this work, we present a general, flexible and harmonious framework to construct interpretable functions in regression analysis with a focus on continuous outcomes. We formulate a functional skeleton in light of users' expectations of interpretability. A new measure based on Mallows's $C_p$-statistic is proposed for model selection to balance approximation, generalizability, and interpretability. We apply this approach to derive a sample size formula in adaptive clinical trial designs to demonstrate the general workflow, and to explain operating characteristics in a Bayesian Go/No-Go paradigm to show the potential advantages of using meaningful intermediate variables. Generalization to categorical outcomes is illustrated in an example of hypothesis testing based on Fisher's exact test. A real data analysis of NHANES (National Health and Nutrition Examination Survey) is conducted to investigate relationships between some important laboratory measurements. We also discuss some extensions of this method. 

\end{abstract}

%



\maketitle


%

\section{Introduction}

Interpretability has always been a desired property for statistical models or machine learning techniques. An interpretable model or method is more appealing in practice for several reasons: first, it is more reliable to outliers or adversarial examples \citep{nguyen2015deep, zhang2021survey}; second, it provides better explanations to be understood by stakeholders, or a better understanding of the underlying scientific questions \citep{lecun2015deep, zhang2021survey}. However, interpretability is a subjective concept, and even its definition is diverse \citep{lipton2018mythos, zhang2021survey}. Many previous works in this area focus on desired properties of interpretable methods \citep{lipton2018mythos}, such as a prerequisite to gain trust \citep{ridgeway1998interpretable, ribeiro2016should}, inferring causal structure \citep{rani2006empirical, pearl2009causality}, fair and ethical consideration \citep{goodman2017european}. 

One popular definition of interpretability is the ability to provide explanations in understandable terms to a human \citep{doshi2017towards, zhang2021survey}. This human-centered definition makes it even harder to consistently evaluate interpretability. For example, a study team may find Deep Neural Networks (DNN; e.g., \citet{goodfellow2016deep}) interpretable based on its multi-layer decomposition of an objective function. However, others may treat DNN as a black-box algorithm with poor interpretation, because the impacts of covariates on the outcome are not as straightforward as those of linear models. Moreover, model complexity plays an important role in interpretability. For example, a deep decision tree \citep{rt} with hundreds of nodes may not be preferred in practice, even though a decision tree is usually treated as a method with good interpretation \citep{craven1995extracting, wu2018beyond}. Additionally, in the context of regression analysis and other problems in supervised learning, estimation accuracy and generalizability should also be accommodated. There is usually a trade-off between accuracy and interpretability \citep{lundberg2017unified}. 

With the above considerations, we propose a general, flexible and harmonious framework to construct interpretable functions. We focus on regression analysis with a continuous outcome to demonstrate our method. Our unified approach has several appealing features: (1) \textbf{General}: this general multi-layer model formulation includes some existing models or interpretation methods as special cases, such as Deep Neural Networks (e.g., \citet{goodfellow2016deep}), and a class of additive feature importance methods in the SHAP framework \citep{lundberg2017unified}. (2) \textbf{Flexible}: the multi-layer skeleton and sets of base interpretable functions can be adjusted based on users' expectations of interpretability. Meaningful intermediate variables can be leveraged to potentially reduce model complexity. (3) \textbf{Harmonious}: we modify Mallows's $C_p$-statistic for model selection to balance model fitting and generalizability, while accommodating model complexity. The final selected method has a good balance between accuracy and interpretability. 

For applications in biomedical-related fields or specifically in the pharmaceutical industry, such interpretable models have great potential to facilitate communication with team members, increase the transparency of decision-making, and give sparks to future scientific questions for investigation. Some motivating examples are: deriving a sample size formula for complex innovative clinical trials in Section \ref{sec:sim1}, streamlining the evaluation of operating characteristics in the Go/No-Go decision-making in Section \ref{sec:sim2}, facilitating hypothesis testing based on Fisher's exact test in Section \ref{sec:sim3}, and investigating relationships between several important laboratory measurements based on a real-world data in Section \ref{sec:real}. 

The remainder of this article is organized as follows. In Section \ref{sec:framework}, we propose this framework to construct interpretable functions. Section \ref{sec:method} provides more details on method implementation. In Section \ref{sec:sim}, we apply this method to two numerical studies, with a specific focus on the general workflow in the first study in Section \ref{sec:sim1}, and the potential advantages of using meaningful intermediate variables in Section \ref{sec:sim2}. Generalization to a categorical outcome is illustrated in Section \ref{sec:sim3}. We conduct a real data analysis of NHANES (National Health and Nutrition Examination Survey) in Section \ref{sec:real}. Discussions are provided in Section \ref{sec:disc}. 

\section{A Framework to Construct Interpretable Functions}
\label{sec:framework}

Denote the unknown function as $y = G(\boldsymbol{x})$, where $\boldsymbol{x}$ is a $M$-dimensional vector, and $y$ is a scalar. In this article, we mainly focus on $y$ as a continuous outcome. Section \ref{sec:sim3} considers a categorical outcome. The outcome $y$ can be generalized to a vector or other generalized outcomes, such as images and time series. 

Our objective is to approximate $G(\boldsymbol{x})$ by $F(\boldsymbol{x}; \boldsymbol{\theta})$ as an interpretable function with parameters $\boldsymbol{\theta}$. We propose to define interpretable functions in regression analysis as multi-layer parametric functions with base components coming from domain knowledge of a specific task. This is motivated by \citet{zhang2021survey}, where interpretability is defined as the ability to provide explanations in understandable terms to a human. Next, we provide further illustration of our proposed definition, and discuss its contributions to interpretation. 

We represent $F(\boldsymbol{x}; \boldsymbol{\theta})$ by $y^\prime$, i.e., $y^\prime \equiv F(\boldsymbol{x}; \boldsymbol{\theta})$. This $F(\boldsymbol{x}; \boldsymbol{\theta})$ is recursively aggregated by $L$ layers of functions. For illustration, $L$ is considered to be $2$. A deeper function can be formulated with this framework for $L>2$. 

In the first layer, we decompose $F(\boldsymbol{x}; \boldsymbol{\theta})$ or $y^\prime$ as
\begin{equation}
\label{equ:f1}
y^\prime = f_1(\boldsymbol{x}_1; \boldsymbol{\theta}_1),
\end{equation}
where $\boldsymbol{x}_1 = (x_{1, 1}, ..., x_{1, J})^T$ is a vector of length $J$, $\boldsymbol{\theta}_1$ is the associated parameter vector and $f_1 \in \mathbb{F}_1$. The set $\mathbb{F}_1$ contains several base functions.  

In the second layer, for $j = 1, \cdots, J$, we write each element $x_{1, j}$ in $\boldsymbol{x}_1$ as
\begin{equation}
\label{equ:f2}
{x}_{1, j} = f_{2, j}(\boldsymbol{x}_{2, j}; \boldsymbol{\theta}_{2, j}),
\end{equation}
where $\boldsymbol{x}_{2, j}$, $\boldsymbol{\theta}_{2, j}$ and $f_{2, j} \in \mathbb{F}_{2, j}$ are analog notations of $\boldsymbol{x}_1$, $\boldsymbol{\theta}_1$ and $f_1$ in the first layer. Since $L=2$ in this motivating framework, we can set $\boldsymbol{x}_{2, j}$ equal to, or as a subset of $\boldsymbol{x}$, which is the original input in $F(\boldsymbol{x}; \boldsymbol{\theta})$. 

With this setup, we break down $F(\boldsymbol{x}; \boldsymbol{\theta})$ by the first layer function $f_1(\boldsymbol{x}_1; \boldsymbol{\theta}_1)$ and the second layer functions $f_{2, j}(\boldsymbol{x}_{2, j}; \boldsymbol{\theta}_{2, j})$, $j = 1, \cdots, J$. The parameter vector $\boldsymbol{\theta} = \left(\boldsymbol{\theta}_1, \boldsymbol{\theta}_{2, 1}, \cdots, \boldsymbol{\theta}_{2, J} \right)^T$ is a stack of all parameters. Given specific functional forms of $f_1  \in \mathbb{F}_1$, and $f_{2, j} \in \mathbb{F}_{2, j}$, $j = 1, ... J$, we can represent $F(\boldsymbol{x}; \boldsymbol{\theta})$ by $F_k \in \mathbb{F}_0$, where the set $\mathbb{F}_0$ has $K$ candidate forms for $F$. 

We consider this formulation of successive layers to contribute to a better interpretation. The outcome is first explained by $\boldsymbol{x}_1$ with a base function $f_1$ in (\ref{equ:f1}), and then $\boldsymbol{x}_1$ are further represented by several second layer functions in (\ref{equ:f2}). This hierarchical representation simplifies the original complex problem by several sub-problems with better interpretations. This construction idea comes from the architecture of Deep Neural Networks (DNN, e.g., \citet{goodfellow2016deep}). A main difference is that base functions $f_1$ in (\ref{equ:f1}) and $f_{2, j}$ in (\ref{equ:f2}) usually have more complex and flexible functional forms than standard activation functions with linear forms of inputs in DNN. Our framework can incorporate more flexible functions to facilitate interpretation. Another perspective is to treat those base functions or components as latent factors of an interpretable function. This framework is also flexible to include several existing models or methods as special cases, for example, some additive methods in the SHAP framework \citep{lundberg2017unified}. 

Specific choices of candidate or base functions in $\mathbb{F}_1$, and $\mathbb{F}_{2, j}$ for $j = 1, ... J$ can be tailored for different problems. They should be based on domain knowledge of a specific task to facilitate interpretation of $F(\boldsymbol{x}; \boldsymbol{\theta})$. We can use equations to connect summary statistics of data and the model parameters, where the equations follow certain principles from domain knowledge. For example, standardized outcomes ($[x-\mu]/\sigma$) and z-score are usually meaningful to statistical audience, and the kernels of existing sample size formula of continuous endpoints are used for the application in adaptive clinical trials in Section \ref{sec:sim1}. For a particular application, stakeholders should discuss what kinds of base functions and the resulting multi-layer function $F$ are interpretable, and then finalize those sets of base functions. A starting point is to specify $L=2$ with 2 layers, and 3 candidates in each of the $\mathbb{F}_1$, and $\mathbb{F}_{2, j}$ for $j = 1, ... J$. However, simple base functions and fewer layers may come with a cost of increased errors for approximation and generalization. More discussion on this trade-off is provided in the next section. If a desired tolerance of estimation or generalization is not reached, deeper models and/or richer candidate sets can be considered. Identifiability is discussed in Section \ref{sec:disc}. 

To better illustrate our framework, we provide three examples. 

\textbf{Example 1:} Consider a target function $y^\prime = F(\boldsymbol{x}) = \left(a^2 + b^3 \right)^2$ with $\boldsymbol{x} = \left(a, b\right)$. We can use $y^\prime = f_1(\boldsymbol{x}_1) = (x_{1,1} + x_{1,2})^2$ as the first layer function, $x_{1,1} = f_{2,1}(x_{2,1}) = x_{2,1}^2$ with $x_{2,1} =a$ and $x_{1,2} = f_{2,2}(x_{2,2}) = x_{2,2}^3$ with $x_{2,2} = b$ as the two second layer functions.

Suppose the interpretation function sets are $\mathbb{F}_1 = \left\{ (x_{1,1} + x_{1,2})^2 \right\}$ with one element, and $\mathbb{F}_{2,j} = \left\{ \theta_{2} x_{2,j}^2, \theta_{2} x_{2,j}^3 \right\}$, for $j = 1, 2$, each with two elements. We consider the following $K = 6$ candidate forms in the set $\mathbb{F}_0$ for $F$:
\begin{align}
\mathbb{F}_0 = & \big\{ (\theta_{2, 1}a^2 + \theta_{2, 2}b^3)^2, (\theta_{2, 1}a^2 + \theta_{2, 2}b^2)^2, (\theta_{2, 1}a^3 + \theta_{2, 2}b^2)^2, (\theta_{2, 1}a^3 + \theta_{2, 2}b^3)^2, \nonumber \\
& (\theta_{2, 1}a^2 + \theta_{2, 2}a^3)^2, (\theta_{2, 1}b^2 + \theta_{2, 2}b^3)^2 \big\}. \nonumber
\end{align}

\textbf{Example 2:} Linear additive functions are usually considered interpretable. For a target function $y^\prime = F(\boldsymbol{x}) = x_1^2 + 6 x_2^3 + 4 x_3$, one can use an additive function in the first layer, and several candidate base functions in the second layer to represent this. 

\textbf{Example 3:} Deep Neural Networks (DNN) has a strong functional representation with several successive layers \citep{goodfellow2016deep, zhan2022dl}. DNN can be a special case of our framework, with candidate base functions in $\mathbb{F}_1$, and $\mathbb{F}_{2, j}, i = 1, ..., J$ being non-linear activation functions on the top of linear predictors. 

\section{Method Implementation}

\label{sec:method}

In this section, we discuss how to obtain an interpretable function $F(\boldsymbol{x}; \boldsymbol{\theta})$ with the above multi-layer setup to approximate the unknown function $G(\boldsymbol{x})$. 

\subsection{Model Fitting based on Input Data}
\label{sec:sgd}

Suppose $\left\{\boldsymbol{x}^{(i)}, y^{(i)} \right\}_{i=1}^N$ are $N$ sets of samples from the target unknown function $y = G(\boldsymbol{x})$, where $y^{(i)} = G\left\{\boldsymbol{x}^{(i)}\right\} + \epsilon_i$, $E(\epsilon_i) = 0$ and $E(\epsilon_i^2) = \sigma_y^2$. Input data can be directly simulated if the underlying mechanism of generating $G(\boldsymbol{x})$ is known, for example, the sample size formula problem in Section \ref{sec:sim1}; or can be the observed data as in Section \ref{sec:real}. 

For a specific $F_k \in \mathbb{F}_0$, $k = 1, ..., K$, we define $\widehat{\boldsymbol{\theta}}_k$ as an estimator of $\boldsymbol{\theta}$ in $F_k(\boldsymbol{x}; \boldsymbol{\theta})$ by the following equation,
\begin{equation}
\label{equ:f_k}
\widehat{\boldsymbol{\theta}}_k = \argmin_{\boldsymbol{\theta}} \frac{1}{N} \sum_{i=1}^N \left[ y^{(i)} - F_k\left\{\boldsymbol{x}^{(i)}; \boldsymbol{\theta}\right\} \right]^2.
\end{equation}
Since we specifically consider $y$ as a continuous outcome, we use the mean squared error (MSE) loss to measure the difference between $\left\{y^{(i)} \right\}_{i=1}^N$ and $\left[F_k\left\{\boldsymbol{x}^{(i)}; {\boldsymbol{\theta}}\right\} \right] _{i=1}^N$. Equation (\ref{equ:f_k}) can be solved by the standard Gradient Descent Algorithm \citep{hastie2009elements}.  

When there are $K$ candidate functional forms in $\mathbb{F}_0$, it becomes a model selection problem to balance approximation, generalization and interpretation. To guide this process, we review Mallows's $C_p$-statistic in the next section, and propose a modified Mallows's $C_p$-statistic ($MC_p$) in Section \ref{sec:mcp}. 

\subsection{Review of the Mallows's $C_p$-statistic}
\label{sec:cp}
	
The Mallows's $C_p$-statistic \citep{mallow1964} is commonly used for model selection in linear regression problems. It is defined as 
\begin{equation}
\label{equ:cp}
C_p(k) = \frac{\text{SSE}_k}{\widehat{\sigma}^2} - N + 2p_k,
\end{equation}
where $\text{SSE}_k$ is the error sum of squares from the model $k$ being considered, $\widehat{\sigma}^2$ is an estimate of the error variance, $N$ is number of samples, and $p_k$ is number of parameters of the model $k$. The MSE of the full model is usually used to calculate $\widehat{\sigma}^2$ \citep{gilmour1996interpretation}. For regression problems under some regularity conditions, $C_p$ has an expected value of $p_k$ \citep{gilmour1996interpretation}. A model with smaller $C_p$ is preferred with smaller $\text{MSE}_k$ for better model fitting, and fewer parameters $p_k$ in (\ref{equ:cp}) for model parsimony. 

\subsection{Modified Mallows's $C_p$-statistic}
\label{sec:mcp}

As a generalization beyond linear regression, we consider a modified Mallows's $C_p$-statistic of model $k$:
\begin{equation}
\label{equ:mcp}
MC(k) = \frac{\text{MSE}_k}{\text{MSE}_{full}} - 1 + \lambda r_k,
\end{equation}
where $\text{MSE}_k$ is the MSE of model $k$, $r_k$ is a measure of model complexity, and $\text{MSE}_{full}$ is computed by the MSE of a relatively deep and wide DNN. This saturated model is denoted as the complex DNN. For the complexity measure $r_k$, one can use the total number of parameters as in $C_p$, the number of layers, or the width of layers \citep{hu2021model}. Most numerical studies in this article use the number of parameters for $r_k$, but Section \ref{sec:sim3} defines $r_k$ as the average number of parameters per layer to provide another perspective of model complexity for multi-layer functions. 

To evaluate approximation and generalization abilities of model $k$, we then define $D$-fold cross-validation $MC_{cv}(k)$ based on training data and $MC^\prime_{cv}(k)$ based on validation data as
\begin{align}
	MC_{cv}(k) & = \frac{\text{MSE}_{cv}(k)}{\text{MSE}_{cv}(f)} - 1 + \lambda r_k, \label{equ:mcp_cv_train} \\
	MC^\prime_{cv}(k) & = \frac{\text{MSE}^\prime_{cv}(k)}{\text{MSE}^\prime_{cv}(f)} - 1 + \lambda r_k, \label{equ:mcp_cv_val} 
\end{align}
where cross-validation MSEs for a model $k$, $k = 1, ..., K$, are defined as,
\begin{align}
	\text{MSE}_{cv}(k) & = \frac{1}{D} \sum_{d=1}^D \text{MSE}_{k,d}, \label{equ:mse_cv_train} \\
	\text{MSE}^\prime_{cv}(k) & = \frac{1}{D} \sum_{d=1}^D \text{MSE}_{k,d}^\prime, \label{equ:mse_cv_val} 
\end{align}
and cross-validation MSEs for the complex DNN or the full model are defined as,
\begin{align}
\text{MSE}_{cv}(f) & = \frac{1}{D} \sum_{d=1}^D \text{MSE}_{full,d}, \label{equ:mse_cv_train_full} \\
\text{MSE}^\prime_{cv}(f) & = \frac{1}{D} \sum_{d=1}^D \text{MSE}_{full,d}^\prime. \label{equ:mse_cv_val_full} 
\end{align}


Then we explain how to compute $\text{MSE}_{k,d}$ in (\ref{equ:mse_cv_train}), $\text{MSE}_{full, d}$ in (\ref{equ:mse_cv_train_full}), $\text{MSE}^\prime_{k,d}$ in (\ref{equ:mse_cv_val}) and $\text{MSE}^\prime_{full, d}$ in (\ref{equ:mse_cv_val_full}). In D-fold cross-validation with a total sample size of $N$, we randomly split data into $D$ portions with $N/D$ samples in each portion. If $N/D$ is not an integer, some portions can have a larger sample size to make sure that each of the $N$ samples is presented and only presented in one of the $D$ portions. For a specific index $d$, for $d = 1, ..., D$, we treat this portion of data as independent validation data. The remaining $D-1$ portions are used as training data to fit model $k$ and the complex DNN. Their training MSEs $\text{MSE}_{k,d}$ and $\text{MSE}_{full, d}$ are calculated. Based on the trained models, validation MSEs $\text{MSE}^\prime_{k,d}$ and $\text{MSE}^\prime_{full, d}$ are computed based on that portion of data for validation. The cross-validated $MC_{cv}(k)$ in (\ref{equ:mcp_cv_train}) and $MC^\prime_{cv}(k)$ in (\ref{equ:mcp_cv_val}) can be obtained after $D$ rounds of calculation. 

For model selections based on either $MC_{cv}(k)$ or $MC^\prime_{cv}(k)$, their final choices can be quite different. In the next section, we propose a method to choose the value of $\lambda$ in (\ref{equ:mcp}) to harmonize selected models based on those two measures. 

Here are some additional remarks on this modified Mallows's $C_p$-statistic:

1. When $\lambda = 2/N$, $r_k$ is the number of parameters and all candidate models have the same $N$, both $C_p$ in (\ref{equ:cp}) and $MC$ in (\ref{equ:mcp}) select the same model because $C_p = N \times MC$. If $\lambda = 0$, then the model selection is purely based on MSE. 

2. In the framework of Section \ref{sec:framework}, one usually evaluates several models with the same $N$, and therefore $MC$ in (\ref{equ:mcp}) does not directly accommodate $N$. The tuning parameter $\lambda$ can be fine-tuned for different $N$.

\subsection{The choice of $\lambda$}
\label{sec:opt}

We propose to choose $\lambda$ in (\ref{equ:mcp}) at a value $\lambda_{opt}$ maximizing correlations between $\left\{MC_{cv}(k) \right\}_{k=1}^K$ and $\left\{MC_{cv}^\prime(k) \right\}_{k=1}^K$. This $\lambda_{opt}$ harmonizes the model selection based on $MC_{cv}$ for model fitting, and $MC_{cv}^\prime$ for model generalization. Grid search method is used to find $\lambda_{opt}$. 

\subsection{Final Model Selection}
\label{sec:final_select}

Two measures $\left\{MC_{cv}(k) \right\}_{k=1}^K$ and $\left\{MC^\prime_{cv}(k) \right\}_{k=1}^K$ are updated based on $\lambda_{opt}$. We recommend selecting the final model with the smallest $MC^\prime_{cv}$, which evaluates model performance based on validation datasets. Parameters $\boldsymbol{\theta}$ in this final model are estimated based on methods described in Section \ref{sec:sgd} with all data.  

\section{Numerical Studies}
\label{sec:sim}

\subsection{Sample Size Formula for Adaptive Clinical Trial Designs}
\label{sec:sim1}

In this section, we utilize our framework to derive a sample size formula for adaptive clinical trial designs. Consider a two-stage two-group adaptive clinical trial with a continuous endpoint, and sample size adaptation in the interim analysis. Denote the sample size per group as $n_1$ in the first stage, and $n_2$ in the second stage. The continuous response $x_{s, j}^{(u)}$ follows a Normal distribution
\begin{equation*}
x_{s, j}^{(u)} \sim \mathcal{N} \left( \mu_j, \sigma^2_j \right),
\end{equation*}
with mean $\mu_j$ and standard deviation $\sigma_j$, for $j=\mathrm{t}$ for treatment and $j=\mathrm{p}$ for placebo. The subscript $s=1, 2$ is the stage index, and the superscript $u = 1, ..., n_s$ is the subject index for a group in stage $s$. Define $\mu_0 = \mu_t - \mu_p$ as the true unknown treatment effect. 

After observing the first stage data $\left\{x_{1, p}^{(u)} \right\}_{u=1}^{n_1}$ in placebo and $\left\{x_{1, t}^{(u)} \right\}_{u=1}^{n_1}$ in treatment, we adjust $n_2$ in the second stage based on the following rule:
\begin{equation}
\label{equ:adp}
n_2 =
\begin{cases}
0.5 \times n_1 & \text{if } \Delta_1 \geq \delta \\
2 \times n_1 & \text{if } \Delta_1 < \delta,
\end{cases}       
\end{equation}
where $\Delta_1 = \sum_{u=1}^{n_1} x_{1, t}^{(u)} / n_1 - \sum_{u=1}^{n_1} x_{1, p}^{(u)} / n_1 $ is the treatment effect estimate in stage 1, and $\delta$ is a clinically meaningful threshold. The $n_2$ is decreased to $0.5 \times n_1$ (rounded to an integer) if a more promising $\Delta_1$ is observed, while increased to $2 \times n_1$ otherwise. Based on observed data from both stages, the final hypothesis testing on the treatment effect $\mu_0$ is based on an inverse normal combinational test with equal weights \citep{bretz2009adaptive, bauer1994evaluation} and one-sided $t$-test to protect Type I error rate at the nominal level $\alpha$. Define the corresponding power as $1-\beta$, where $\beta$ is the Type II error rate. 

For illustration, we consider a setting to learn the sample size formula $F(\boldsymbol{x})$ with $\boldsymbol{x} = (\mu_0, \alpha, \beta)$ to compute $n_1$, given known $\sigma_p = \sigma_t = 1$, and $\delta = 0.3$. For a specific adaptive design in (\ref{equ:adp}), the total sample size of both groups from two stages is either $6\times n_1$ or $3\times n_1$ depending on observed $\Delta_1$ in stage 1. 

Following the framework in Section \ref{sec:method}, we simulate a total of $N = 1,000$ sets of data. For the input $\boldsymbol{x} = (\mu_0, \alpha, \beta)$ and the output $n_1$, we sample $\mu_0$ from the Uniform distribution $Unif(0.1, 0.6)$, $\alpha$ from $Unif(0.01, 0.15)$, and $n_1$ from a uniform distribution on integers from $10$ to $60$. We transform the outcome as $n_1/60$ to be bounded between $0$ and $1$. Given specific $\mu_0$, $\alpha$ and $n_1$, the corresponding $\beta$ (1-power) is calculated based on $10^4$ Monte Carlo samples. Ranges of those generating distributions can be adjusted.  

Suppose that the study team wants to construct an interpretable function $F(\boldsymbol{x}; \boldsymbol{\theta})$ with $L=2$ layers. In the first layer function of (\ref{equ:f1}), the input variable $\boldsymbol{x}_1 = (x_{1, 1}, x_{1, 2})$ has $J=2$ elements. The candidate function set $\mathbb{F}_1$ is considered with 3 potential forms,
\begin{align} 
\mathbb{F}_1 = \big\{ & f_1^{(1)} = \theta_1 \left(x_{1, 1}^2 + x_{1, 2}^2\right), \nonumber \\
& f_1^{(2)} = \theta_1 x_{1, 1} + \theta_2 x_{1, 2}, \nonumber \\
& f_1^{(3)} = \theta_1 x_{1, 1} + \theta_2 x_{1, 2} + \theta_3 \big\} \nonumber
\end{align}
These base functions $ f_1^{(1)}$, $ f_1^{(2)}$ and $ f_1^{(3)}$ have one, two and three parameter(s), respectively. Their linear forms are usually helpful for interpretation. For $j=1, 2$, the input of the first layer function ${x}_{1, j}$ is further used as the output of the second layer function in (\ref{equ:f2}). The corresponding input $\boldsymbol{x}_{2, j}$ has all three elements in $\boldsymbol{x} = (\mu_0, \alpha, \beta)$. The set $\mathbb{F}_{2, j}$ of $f_{2, j}$ has four candidates,
\begin{align} 
\mathbb{F}_{2, j} = \big[ & f_{2}^{(1)} =  \left\{ \left( Z_{1-\alpha} + Z_{1-\beta} \right) / \mu_0 \right\}^2 + \theta_1, \nonumber \\
& f_{2}^{(2)} = \theta_1\left\{ \left( Z_{1-\alpha} + Z_{1-\beta} + \theta_2 \right) / \mu_0 \right\}^2, \nonumber \\
& f_{2}^{(3)} =\theta_1\left\{ \left( Z_{1-\alpha} + Z_{1-\beta} + \theta_2 \right) / \mu_0 + \theta_3 \right\}^2 , \nonumber \\
& f_{2}^{(4)} =\left\{ \left( \theta_1 Z_{1-\alpha} + \theta_2 Z_{1-\beta} \right) / \mu_0 + \theta_3 \right\}^2 + \theta_4 \big], \nonumber 
\end{align}
where $Z_{1-\alpha}$ is the upper $\alpha$ quantile of the standard Normal distribution. Their functional forms are based on the kernel of the sample size formula of the two-sample continuous outcome, but with different numbers of additional free parameters $\theta$. 

In total, we consider $K = 18$ functional forms of $F$ in the set $\mathbb{F}_0$, as listed in Table \ref{tab:sim_1}. We set the complexity measure $r$ in (\ref{equ:mcp}) as the total number of parameters of a model, which is shown in the column "r" of  Table \ref{tab:sim_1}. For example $r=4$ in $f_1^{(1)}\left\{ f_{2}^{(1)}, f_{2}^{(2)} \right\}$. To calculate $\text{MSE}_{full}$ in equation (\ref{equ:mcp}), we use a relatively complex DNN with two hidden layers and $60$ nodes per layer. Other training parameters are set with a dropout rate of $0.2$, a total of $100$ epochs, a batch size of $10$, a learning rate of $0.005$, ReLU as the activation function for hidden layers, and sigmoid as the last-layer activation function. Additionally, we evaluate a benchmark model as a relatively simple DNN with one hidden layer and $2$ nodes in that layer. A 5-fold cross-validation with $D=5$ is implemented to compute cross-validation MSEs. The above settings are used in later numerical studies, unless specified otherwise. 

\begin{figure}[h]
	\centering
	\includegraphics[width=16cm]{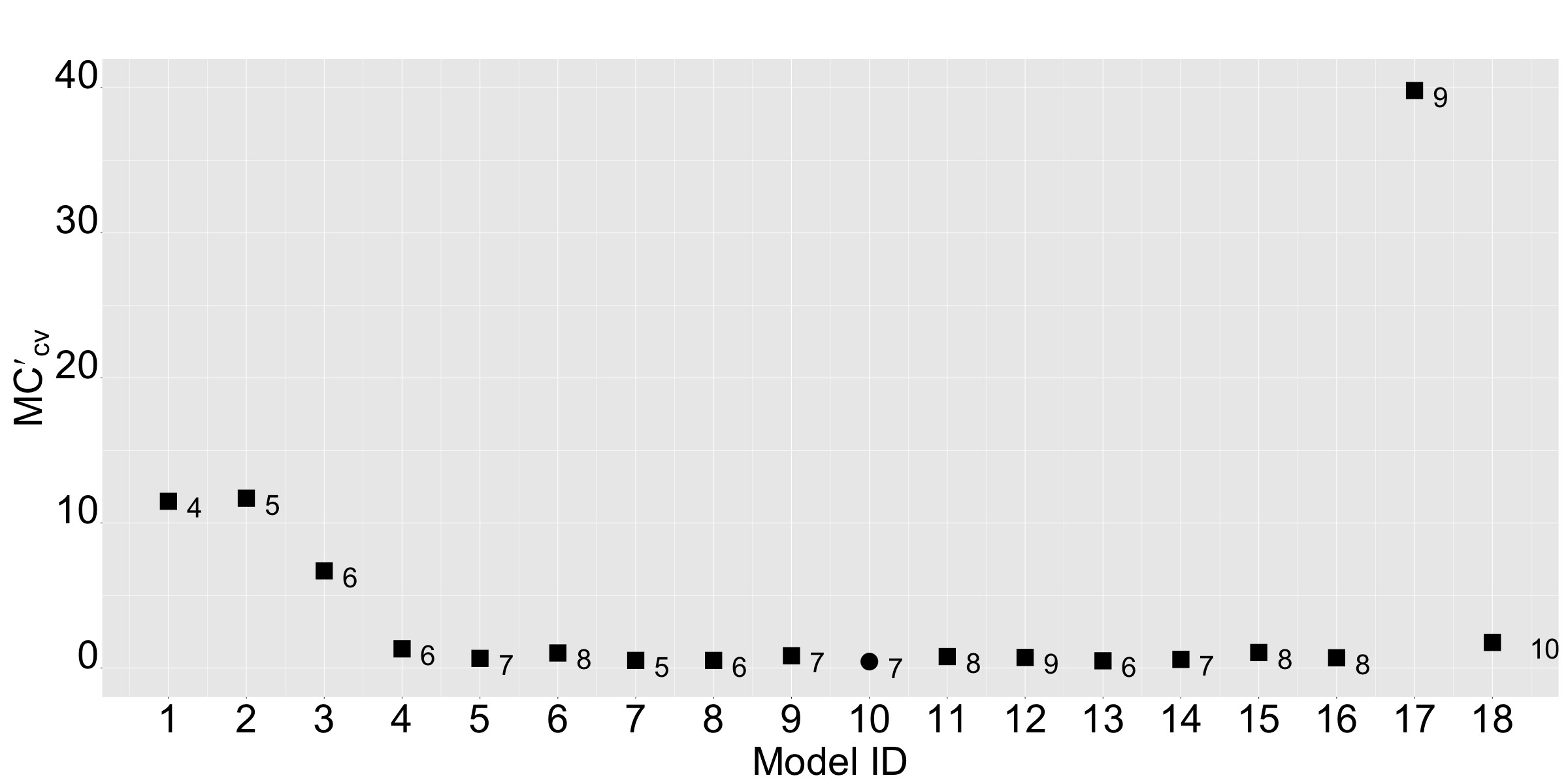}
	\caption{Visualization of $MC^\prime_{cv}$ from $K=18$ candidate functions. The number of parameters is displayed on the right-hand side of each model. Model 10 with the smallest $MC^\prime_{cv}$ is in a circle, while the rest of the models are labeled by squares. }
	\label{fig:sim_1}
\end{figure}

Among those $K=18$ functions, Model 12 has the smallest training $\text{MSE}_{cv}$ in cross-validation, but Model 10 has the smallest validation $\text{MSE}^\prime_{cv}$ in cross-validation (Table \ref{tab:sim_1}). To harmonize these two choices and incorporate model complexity, we further compute $MC_{cv}$ and $MC^\prime_{cv}$ as discussed in Section \ref{equ:mcp}. By grid search optimization, $\lambda_{opt}$ is selected at a value of $0.14$. With either $MC_{cv}$ or $MC^\prime_{cv}$, Model 10 can be chosen with the smallest value. As compared to the benchmark simple DNN, this model 10 has much smaller $\text{MSE}_{cv}$ and $\text{MSE}^\prime_{cv}$, and therefore has smaller $MC_{cv}$ and $MC^\prime_{cv}$. The complex DNN has similar $\text{MSE}_{cv}$ and $\text{MSE}^\prime_{cv}$ with model 10, but its $MC_{cv}$ and $MC^\prime_{cv}$ are much higher because of the large number of parameters. Figure \ref{fig:sim_1} visualizes $MC^\prime_{cv}$ of all $18$ candidate functions. 

After cross-validation to select Model 10 as the final one, we use all $N = 1,000$ data to obtain estimated parameters. This model has the following functional form to represent the sample size $n_1$ by the input $\boldsymbol{x} = (\mu_0, \alpha, \beta)$:
\begin{equation*}
n_1 = 0.056 \left\{ \left( Z_{1-\alpha} + Z_{1-\beta} + 1.64 \right) / \mu_0 \right\}^2  + 0.758 \left\{ \left( Z_{1-\alpha} + Z_{1-\beta} - 0.49 \right) / \mu_0 + 1.40 \right\}^2  
\end{equation*}
It has $f_1^{(2)}$ with an additive form as the first layer function, and $f_2^{(2)}$ and $f_2^{(3)}$ as two second layer functions with kernels from the sample size formula of the continuous outcome. 


\begin{table}[ht]
\centering
\begin{tabular}{ccccccc}
\hline
Model ID & Functional Form & $r$ & $\text{MSE}_{cv}$ & $\text{MSE}_{cv}^\prime$  & $MC_{cv}$ & $MC_{cv}^\prime$ \\
\hline
1 & $f_1^{(1)}\left\{ f_{2}^{(1)}, f_{2}^{(2)} \right\}$ & 4 & 0.05452 & 0.05503 & 13.6 & 11.5 \\
2 & $f_1^{(1)}\left\{ f_{2}^{(1)}, f_{2}^{(3)} \right\}$ & 5 & 0.05468 & 0.05534 & 13.7 & 11.7 \\
3 & $f_1^{(1)}\left\{ f_{2}^{(1)}, f_{2}^{(4)} \right\}$ & 6 & 0.02960 & 0.03162 & 7.4 & 6.7 \\
4 & $f_1^{(1)}\left\{ f_{2}^{(2)}, f_{2}^{(3)} \right\}$ & 6 & 0.00667 & 0.00680 & 1.6 & 1.3 \\
5 & $f_1^{(1)}\left\{ f_{2}^{(2)}, f_{2}^{(4)} \right\}$ & 7 & 0.00301 & 0.00313 & 0.8 & 0.7 \\
6 & $f_1^{(1)}\left\{ f_{2}^{(3)}, f_{2}^{(4)} \right\}$ & 8 & 0.00397 & 0.00420 & 1.1 & 1.0 \\
7 & $f_1^{(2)}\left\{ f_{2}^{(1)}, f_{2}^{(2)} \right\}$ & 5 & 0.00379 & 0.00378 & 0.7 & 0.5 \\
8 & $f_1^{(2)}\left\{ f_{2}^{(1)}, f_{2}^{(3)} \right\}$ & 6 & 0.00315 & 0.00312 & 0.6 & 0.5 \\
9 & $f_1^{(2)}\left\{ f_{2}^{(1)}, f_{2}^{(4)} \right\}$ & 7 & 0.00394 & 0.00398 & 1.0 & 0.8 \\
10 & $f_1^{(2)}\left\{ f_{2}^{(2)}, f_{2}^{(3)} \right\}$ & 7 & 0.00208 & \underline{0.00211} & \underline{0.5} & \underline{0.4} \\
11 & $f_1^{(2)}\left\{ f_{2}^{(2)}, f_{2}^{(4)} \right\}$ & 8 & 0.00326 & 0.00305 & 1.0 & 0.8 \\
12 & $f_1^{(2)}\left\{ f_{2}^{(3)}, f_{2}^{(4)} \right\}$ & 9 & \underline{0.00207} & 0.00215 & 0.8 & 0.7 \\
13 & $f_1^{(3)}\left\{ f_{2}^{(1)}, f_{2}^{(2)} \right\}$ & 6 & 0.00290 & 0.00301 & 0.6 & 0.5 \\
14 & $f_1^{(3)}\left\{ f_{2}^{(1)}, f_{2}^{(3)} \right\}$ & 7 & 0.00281 & 0.00281 & 0.7 & 0.6 \\
15 & $f_1^{(3)}\left\{ f_{2}^{(1)}, f_{2}^{(4)} \right\}$ & 8 & 0.00417 & 0.00434 & 1.2 & 1.1 \\
16 & $f_1^{(3)}\left\{ f_{2}^{(2)}, f_{2}^{(3)} \right\}$ & 8 & 0.00261 & 0.00268 & 0.8 & 0.7 \\
17 & $f_1^{(3)}\left\{ f_{2}^{(2)}, f_{2}^{(4)} \right\}$ & 9 & 0.16509 & 0.18238 & 42.7 & 39.8 \\
18 & $f_1^{(3)}\left\{ f_{2}^{(3)}, f_{2}^{(4)} \right\}$ & 10 & 0.00644 & 0.00627 & 2.1 & 1.8 \\
\\
Benchmark & & 11 & 0.04975 & 0.05068 & 13.3 & 11.5 \\
\\
Complex DNN & & 3961 & 0.00390 & 0.00460 & 555.1 & 555.1 \\ 
\hline
\end{tabular}
\caption{Summary of model complexity (number of parameters) $r$, training MSE in cross-validation $\text{MSE}_{cv}$, validation MSE in cross-validation $\text{MSE}^\prime_{cv}$, $MC_{cv}$ based on training data and $MC_{cv}^\prime$ based on validation data of $K=18$ candidate functions, the benchmark model as a simple DNN and the complex DNN. Within each of the 4 columns to the right, the smallest number is underlined. Model 12 has the smallest $\text{MSE}_{cv}$, but Model 10 has the smallest $\text{MSE}^\prime_{cv}$. With either $MC_{cv}$ or $MC_{cv}^\prime$ to harmonize approximation and generalization and to incorporate model complexity, Model 10 has the overall best performance. }
	\label{tab:sim_1}
\end{table}

\subsection{One-sample Binary Endpoint Go/No-Go Evaluation}
\label{sec:sim2}

Go/No-Go (GNG) assessments of the Phase 3 clinical program are not merely based on achieving statistical significance in proceeding Phase 2 studies, but mainly on the confidence to achieve certain clinically meaningful differences or target product profiles (TPPs). In this section, we use our proposed framework to construct interpretable functions to evaluate operating characteristics in GNG assessment based on a Bayesian paradigm in \cite{pulkstenis2017bayesian} with data from a proof-of-concept one-sample Phase 2 clinical trial with a binary endpoint. We mainly focus on the potential advantages of using meaningful intermediate variables as inputs to reduce model complexity. 

Suppose that the number of patients achieving this binary endpoint $n_r$ follows a Binomial distribution,
\begin{equation}
\label{sim2:bin}
n_r \sim Bin (n, q),
\end{equation}
where $n$ is the total number of patients, and $q$ is the response rate. We consider a conjugate setup where $q$ has a prior Beta distribution,
\begin{equation*}
q \sim Beta (q_a, q_b).
\end{equation*}
Define $T_{min}$ as the minimal level of acceptable efficacy, and $T_{base}$ be the desired level of efficacy, with $T_{min} \leq T_{base}$. With observed Phase 2 data $(n_r, n)$, one will make a Go decision of Phase 3 if,  
\begin{align}
Pr(q \geq T_{min}|n_r, n) > \tau_{min} \text{ and } Pr(q \geq T_{base}|n_r, n) > \tau_{base},
\end{align}
where $\tau_{min}$ and $\tau_{base}$ are thresholds between $0$ and $1$.  

At the study design stage, the following expected value of making a Go decision is of particular interest,
\begin{equation}
\label{sim2:y}
y = \sum_{n_r=0}^n I\Big\{Pr(q \geq T_{min}|n_r, n) > \tau_{min} \text{ and } Pr(q \geq T_{base}|n_r, n) > \tau_{base} \Big\} \pi(n_r|q_0, n),
\end{equation}
where $\pi(n_r|q_0, n)$ is the probability mass function of the Binomial distribution in (\ref{sim2:bin}) under a specific rate $q_0$. In practice, this quantity is usually evaluated by Monte Carlo samples given certain design parameters. 

Consider a setting with $\tau_{min} = 0.8$, $\tau_{base} = 0.1$ and $n=40$. We implement our proposed method to construct interpretable functions of $y$ in (\ref{sim2:y}) given $T_{min}$, $T_{base}$ and $q$. To generate training data, $T_{min}$ is simulated from $Unif(0.1, 0.3)$, $T_{base}$ is from $T_{min}+Unif(0.05, 0.2)$, and $q_0$ is from $Unif(0.1, 0.6)$. All other simulation specifications are the same as in the previous section, except that the set $\mathbb{F}_{2, j}$ of $f_{2, j}$ has the following four candidates,
\begin{align} 
\mathbb{F}_{2, j} = \big\{ & f_{2}^{(1)} = \theta_1 \left(x_{2, 1}^2 + x_{2, 2}^2 + x_{2, 3}^2  \right), \nonumber \\
& f_{2}^{(2)} = \theta_1 \left(x_{2, 1} + x_{2, 2} + x_{2, 3} \right) + \theta_2, \nonumber \\
& f_{2}^{(3)} = \theta_1 x_{2, 1} + \theta_2 x_{2, 2} + \theta_3 x_{2, 3}, \nonumber \\
& f_{2}^{(4)} = \theta_1 x_{2, 1} + \theta_2 x_{2, 2} + \theta_3 x_{2, 3} + \theta_4 \big\}, \nonumber 
\end{align}
where $\boldsymbol{x} = (x_{2,1}, x_{2,2}, x_{2,3})$. This set is modified from the sample size formula kernel in the previous section to linear base functions to facilitate interpretation in this specific problem. 

\begin{figure}
	\centering
	\begin{subfigure}[b]{1\textwidth}
		\includegraphics[width=1\linewidth]{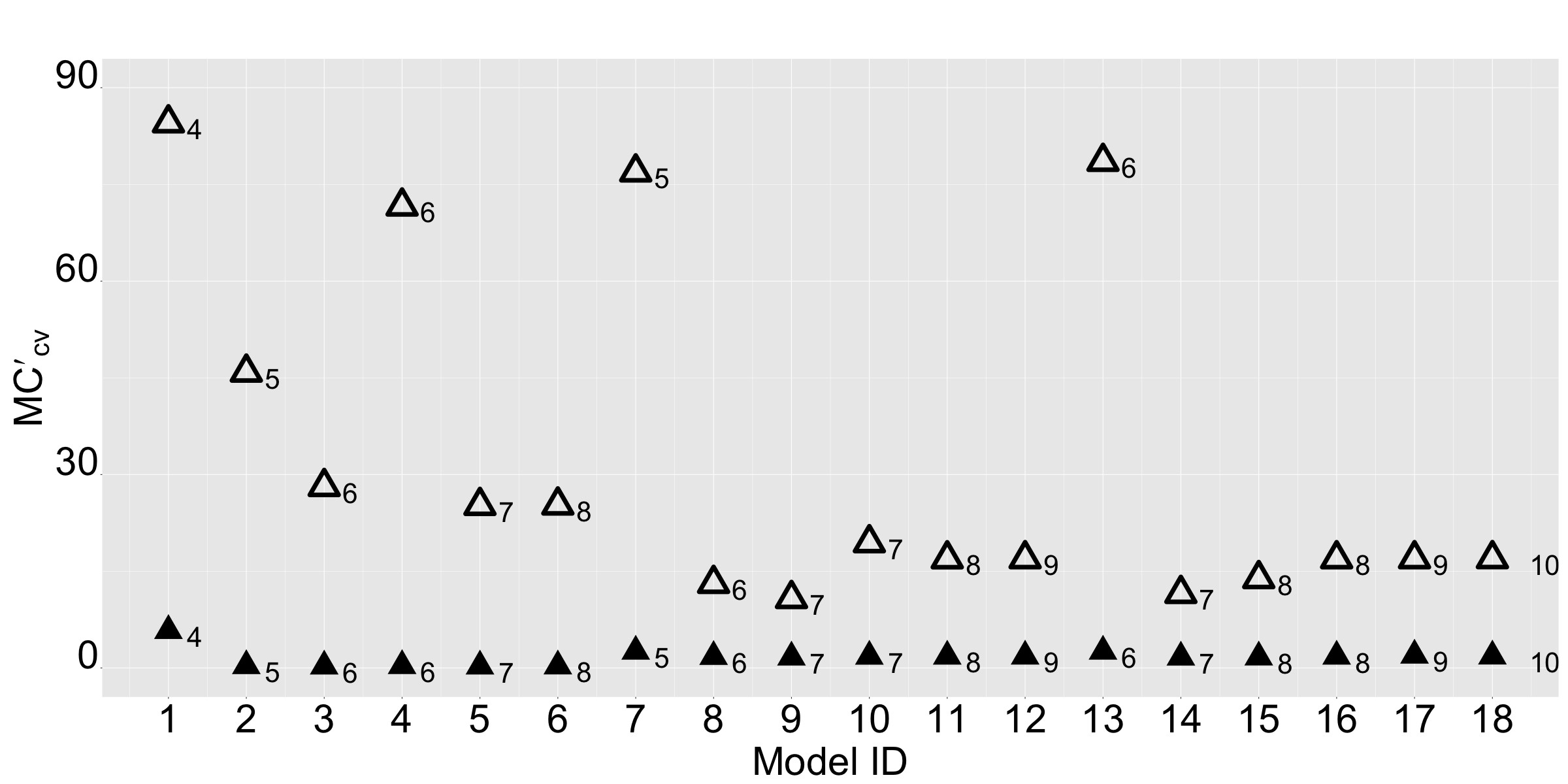}
		\caption{}
		\label{fig:Ng1} 
	\end{subfigure}
	
	\begin{subfigure}[b]{1\textwidth}
		\includegraphics[width=1\linewidth]{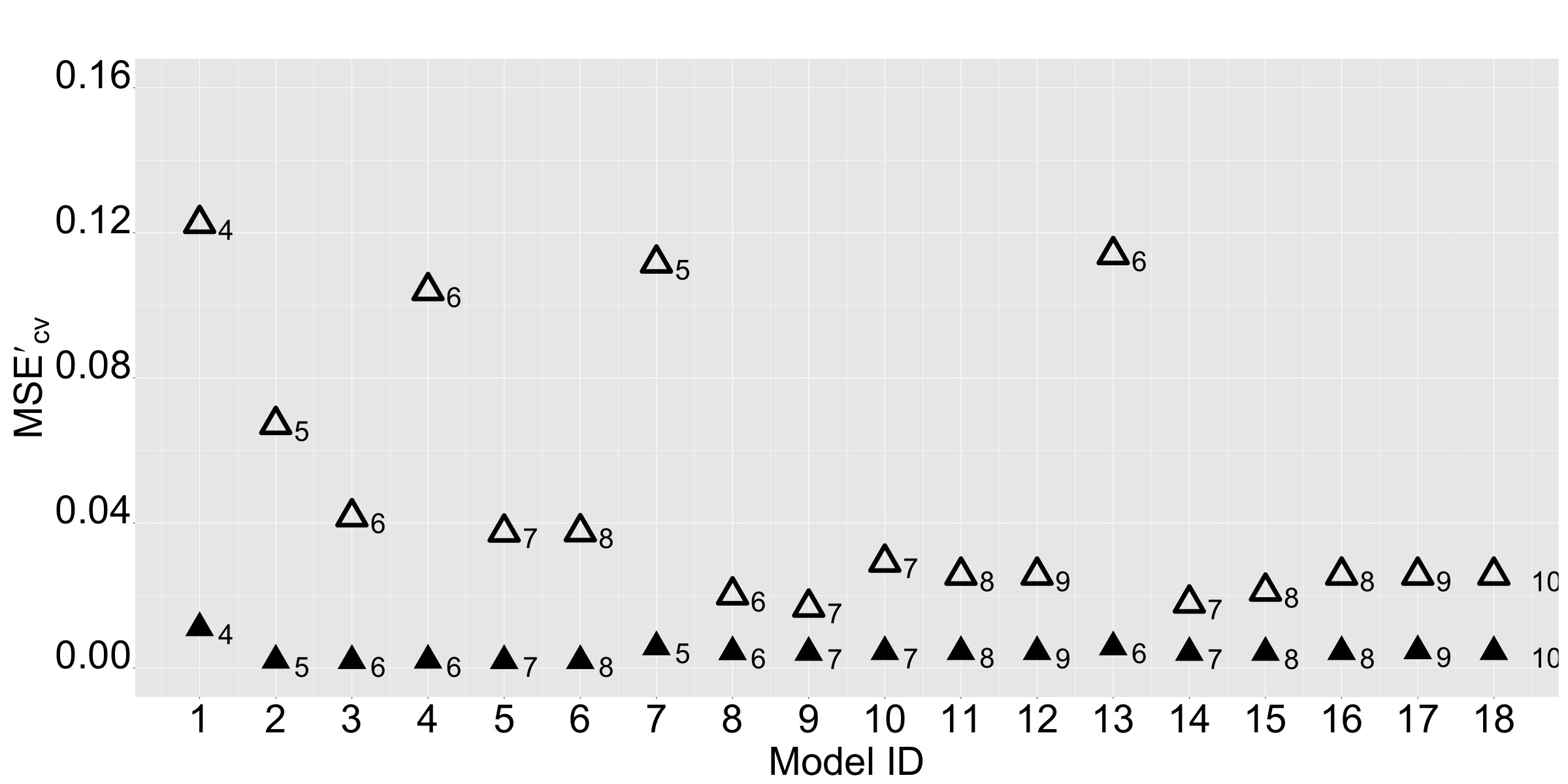}
		\caption{}
		\label{fig:Ng2}
	\end{subfigure}
	\caption[TBD]{Visualization of $MC_{cv}^\prime$ (top figure) and $\text{MSE}^\prime_{cv}$ (bottom figure) of $\boldsymbol{x}_{ori}$ in hollow triangles and $\boldsymbol{x}_{int}$ in solid triangles. The number of parameters is displayed on the right-hand side of each model. Models with $\boldsymbol{x}_{int}$ have smaller $MC_{cv}^\prime$ and $\text{MSE}^\prime_{cv}$ than their corresponding models with $\boldsymbol{x}_{ori}$. }
	\label{fig:int}
\end{figure}

On the input variables $\boldsymbol{x}$ in $F(\boldsymbol{x};\boldsymbol{\theta})$ to predict $y$, we can directly use those original variables $T_{min}$, $T_{base}$ and $q_0$, denoted as $\boldsymbol{x}_{ori} = (T_{min}, T_{base}, q_0)$. Alternatively, we can also use $\boldsymbol{x}_{int} = \Big\{Pr(q \geq T_{min}|n\times q_0, n), Pr(q \geq T_{base}|n\times q_0, n), q_0\Big\}$ by replacing TPPs with their corresponding posterior probabilities in (\ref{sim2:y}). The performance in terms of $MC_{cv}^\prime$ and $\text{MSE}_{cv}^\prime$ of all $K=18$ models with $\boldsymbol{x}_{ori}$ and $\boldsymbol{x}_{int}$ are shown in Figure \ref{fig:int}. Models with $\boldsymbol{x}_{int}$ have a better performance of $MC_{cv}^\prime$ and $\text{MSE}_{cv}^\prime$ than $\boldsymbol{x}_{ori}$. These findings suggest that the model constructed by $\boldsymbol{x}_{int}$ can generally achieve a better $MC_{cv}^\prime$ or $\text{MSE}_{cv}^\prime$, or is simpler with fewer number of parameters under a tolerance level, as compared to $\boldsymbol{x}_{ori}$.

This technique of using meaningful intermediate variables or feature engineering is widely used in other machine learning problems \citep{Chollet, zhan2022finite}. In a problem of reading the time on a clock in \cite{Chollet}, a Convolutional Neural Network (CNN) can be trained to predict time based on raw pixels of the clock image. With a better input of angles of clock hands, we can directly tell the time with rounding and dictionary lookup. However, leveraging intermediate variables may lead to information loss, and may further result in poorer performance than the original variables. Some related discussion on missing data imputation strategies of a dichotomized endpoint based on a continuous endpoint can be found in \cite{floden2019imputation, li2022analyzing}. Therefore, both original and meaningful intermediate input variables can be evaluated to find a proper input set. 
 
\subsection{Hypothesis Testing base on Fisher's Exact Test}
\label{sec:sim3}

In this section, we demonstrate our method in a two-class classification problem to construct an interpretable function for hypothesis testing with Fisher's exact test. Since the p-value calculation involves a summation of probability mass functions from more extreme data, we aim to construct a function for hypothesis testing from another perspective. 

Suppose that we have two groups of binary data with an equal sample size $n$, and $q_1$ and $q_2$ as the number of responders in each group. The input variable is $\boldsymbol{x} = (q_1, q_2, n)$, and the output label $\boldsymbol{y}$ is a vector of length $2$. We are interested in testing if the odds ratio of group 2 versus group 1 is larger than 1. The output label $\boldsymbol{y}$ is $(0, 1)$ if the corresponding one-sided p-value from Fisher's exact test is less than $0.05$, and is $(1, 0)$, otherwise. Our objective is to construct $F(\boldsymbol{x};\boldsymbol{\theta}) \equiv \boldsymbol{y}^\prime$ with $L = 2$ layers to predict $\boldsymbol{y}$. In (\ref{equ:mcp_cv_train}) and (\ref{equ:mcp_cv_val}), we consider $r_k$ as the average number of parameters per layer, instead of the total number of parameters in other sections. 

In the first layer function to obtain $\boldsymbol{y}^\prime$ based on two inputs $x_{1, 1}$ and $x_{1, 2}$, there are 3 candidate function forms in $\mathbb{F}_1$ (\ref{equ:sim_4_f1}) with a sigmoid kernel to ensure that two elements in $\boldsymbol{y}^\prime$ are bounded within 0 and 1 with a sum of 1. The linear predictor has one parameter in the first candidate form (\ref{equ:sim_4_f1_y1}), two parameters in (\ref{equ:sim_4_f1_y2}) and three parameters in (\ref{equ:sim_4_f1_y3}). 

\begin{align} 
\mathbb{F}_1 = & \left( f_1^{(m)} = \left\{\frac{1}{1+ exp\left[y_{1}^{(m)} \right]}, \frac{exp\left[y_{1}^{(m)} \right]}{1+ exp\left[y_{1}^{(m)} \right]} \right\} , m = 1, 2, 3 \right) \label{equ:sim_4_f1}
\end{align}
\begin{align}
y_{1}^{(1)} = & \theta_1 x_{1,1}+\theta_1 x_{1, 2} \label{equ:sim_4_f1_y1} \\
y_{1}^{(2)} = & \theta_1 x_{1,1}+ \theta_2 x_{1, 2} \label{equ:sim_4_f1_y2} \\
y_{1}^{(3)} = & \theta_1 x_{1,1} + \theta_2 x_{1, 2} + \theta_3 \label{equ:sim_4_f1_y3}
\end{align}

The output of the second layer function is the input of the first layer function $x_{1, j}$ ($j = 1, 2$). The input of the second layer function is $\boldsymbol{x} = (q_1, q_2, n)$. The three candidate forms in $\mathbb{F}_{2, j}$ have one, two and three parameters, respectively: 
\begin{align} 
\mathbb{F}_{2, j} = \Bigl\{ & f_{2}^{(1)} = \frac{\theta_1 q_1 + \theta_1 q_2}{n}, \nonumber \\
& f_{2}^{(2)} = \frac{\theta_1 q_1 + \theta_2 q_2}{n}, \nonumber \\
& f_{2}^{(3)} = \frac{\theta_1 q_1 + \theta_2 q_2}{n+\theta_3} \Bigl\}. \nonumber 
\end{align}
They have base components of $q_1/n$ and $q_2/n$ interpreted as observed response rates in each group.  

To implement our proposed method with categorical outcome $\boldsymbol{y}$, we use cross-entropy instead of MSE as the loss function in Section \ref{sec:mcp}. "$CE$" is used in notations to replace "$MSE$". When training the complex DNN and the benchmark simple DNN, we follow settings in Section \ref{sec:sim1}, except that the last-layer activation is softmax for the categorical outcome, and a smaller learning rate at $0.0005$ to stabilize gradient update. The $\lambda_{opt}$ in Section \ref{sec:opt} is computed at $0.12$. The model with $f_1^{(3)}$ as the first layer function, and $f_2^{(1)}$ and $f_2^{(3)}$ as two second layer functions is the selected one with the smallest $MC_{cv}^\prime$. 

Table \ref{tab:sim4} shows performance of the selected model, the benchmark simple DNN and the complex DNN. $A_{cv}$ and $A_{cv}^\prime$ are accuracies for binary classification based on training and validation datasets in cross-validation, respectively. Both the selected model and the complex DNN have small cross-entropy loss and high accuracy. However, the complex DNN has large $MC_{cv}$ and $MC_{cv}^\prime$ incorporating model complexity. On the other hand, the selected model achieves the best $MC$ with a simpler model. It is also superior to the benchmark simple DNN. 

\begin{table}[ht]
\centering
\begin{tabular}{ccccccccc}
\hline
Model & $r$ & $\text{CE}_{cv}$ & $\text{CE}_{cv}^\prime$  & $MC_{cv}$ & $MC_{cv}^\prime$ & $A_{cv}$ & $A_{cv}^\prime$ \\
\hline
Selected & 4 & 0.025 & 0.032 & 0.3 & 0.3 & 98.8\% & 98.7\% \\
\\
Benchmark & 7 & 0.547 & 0.550 & 19.1 & 15.8 & 76.3\% & 76.3\% \\
\\
Complex DNN & 1341 & 0.028 & 0.035 & 161.0 & 161.0 & 98.9\% & 98.6\% \\
\hline
\end{tabular}
\caption{Summary of model complexity (average number of parameters per layer) $r$, training cross-entropy in cross-validation $\text{CE}_{cv}$, validation cross-entropy in cross-validation $\text{CE}^\prime_{cv}$, training $MC_{cv}$, validation $MC_{cv}^\prime$, training accuracy in cross-validation $A_{cv}$, and validation accuracy in cross-validation $A_{cv}^\prime$ of the selected model, the benchmark simple DNN and the complex DNN.}
\label{tab:sim4}
\end{table}

\section{Analysis of NHANES (National Health and Nutrition Examination Survey) }

\label{sec:real}

In this section, we conduct a real data analysis based on the 2017-2018 survey cycle of the National Health and Nutrition Examination Survey (NHANES), which is a program of studies designed to assess the health and nutritional status of adults and children in the United States \citep{NHANES}. 

Suppose our scientific question is to estimate/predict albumin value, which is an important laboratory measurement, based on several other lab values. The outcome is scaled with $(\log(y)+3)/14$ to be bounded within $0$ and $1$. There are $10$ other lab values considered as potential explanatory covariates: creatinine, High-density lipoprotein (HDL) cholesterol, C-reactive protein, insulin, triglyceride, Low-density lipoprotein (LDL) cholesterol, iodine, vitamin C, vitamin D2, vitamin D3. All covariates are standardized with a mean of $0$ and a standard deviation of $1$. A total of $928$ records with complete data are included in this analysis. 

To facilitate interpretation, our method is implemented to construct an interpretable function $F(\boldsymbol{x})$ with $L=2$ layers, with $J=2$ elements in the input $\boldsymbol{x}_1 = (x_{1, 1}, x_{1, 2})$ of the first layer function, and $M=2$ elements in $\boldsymbol{x} = (z_1, z_2)$. There are 3 candidate forms in $\mathbb{F}_1$ for the first layer function,
\begin{align} 
\mathbb{F}_1 = \Big\{ & f_1^{(1)} = \theta_1 \left(x_{1, 1} + x_{1, 2} \right)^2, \nonumber \\
& f_1^{(2)} = \theta_1 x_{1, 1}^3 + \theta_2 x_{1, 2}^3, \nonumber \\
& f_1^{(3)} = \theta_1 x_{1, 1} + \theta_2 x_{1, 2} + \theta_3 \Big\}. \nonumber
\end{align}
For the two second layer functions $j = 1, 2$, $\mathbb{F}_{2, j}$ also has 3 potential forms,
\begin{align} 
\mathbb{F}_{2, j}  = \Big\{ & f_2^{(1)} = \theta_1 \left(z_{1} + z_{2} \right)^2, \nonumber \\
& f_2^{(2)} = \theta_1 z_{1}^3 + \theta_2 z_{2}^3, \nonumber \\
& f_2^{(3)} = \theta_1 z_{1} + \theta_2 z_{2} + \theta_3 \Big\}. \nonumber
\end{align}
Similar to the setup in Section \ref{sec:sim1} and \ref{sec:sim2}, those sets contain base functions with components that are commonly used in this application, such as additive forms, quadratic terms and cubic terms.  

There are $9$ functional forms of constructing $F(\boldsymbol{x}; \boldsymbol{\theta})$ given a specific $\boldsymbol{x}$. With a total of $10$ potential explanatory variables, there are $45$ combinations of choosing $M=2$ variables in $\boldsymbol{x}$. Therefore, there are a total of $K = 405 = 9*45$ elements in the set $\mathbb{F}_0$. 

In our proposed framework, we conduct a 4-fold cross-validation ($D=4$) with $928/4 = 232$ samples in each portion. As in Section \ref{sec:opt}, we compute $\lambda_{opt} = 0.69$. We select the final model with the smallest $MC^\prime_{cv}$ among all $K=405$ models. Table \ref{tab:real} evaluates the performance of this selected model, the benchmark simple DNN, and the complex DNN. In terms of MSE, all 3 models have similar performance, but the complex DNN has the smallest $\text{MSE}_{cv}$ at $0.0041$ and the benchmark simple DNN has the smallest $\text{MSE}_{cv}^\prime$ at $0.0071$. With the proposed $MC$ statistics to balance model complexity, the selected model has the best $MC_{cv}$ and $MC^\prime_{cv}$. 

\begin{table}[ht]
	\centering
	\begin{tabular}{ccccccc}
		\hline
		Model & $r$ & $\text{MSE}_{cv}$ & $\text{MSE}_{cv}^\prime$  & $MC_{cv}$ & $MC_{cv}^\prime$ \\
		\hline
Selected & 5 & 0.0078 & 0.0079 & 4.4 & 3.5 \\
\\
Benchmark & 25 & 0.0067 & 0.0071 & 17.9 & 17.2 \\
\\
Complex DNN & 4381 & 0.0041 & 0.0076 & 3025.9 & 3025.9 \\
		\hline
	\end{tabular}
	\caption{Summary of model complexity (number of parameters) $r$, training MSE in cross-validation $\text{MSE}_{cv}$, validation MSE in cross-validation $\text{MSE}^\prime_{cv}$, training $MC_{cv}$ and validation $MC_{cv}^\prime$ of the selected model, the benchmark simple DNN and the complex DNN.}
	\label{tab:real}
\end{table}

This selected model uses $f_1^{(1)}$ in the first layer, and $f_2^{(1)}$ and $f_2^{(3)}$ in the second layer, with a total of $5$ parameters. The input $\boldsymbol{x}$ has creatinine and HDL cholesterol as two explanatory variables. We then use all $N = 928$ data to conduct the final model fitting.  

\begin{figure}
	\centering
	\begin{subfigure}[b]{0.475\textwidth}
		\includegraphics[width=1\linewidth]{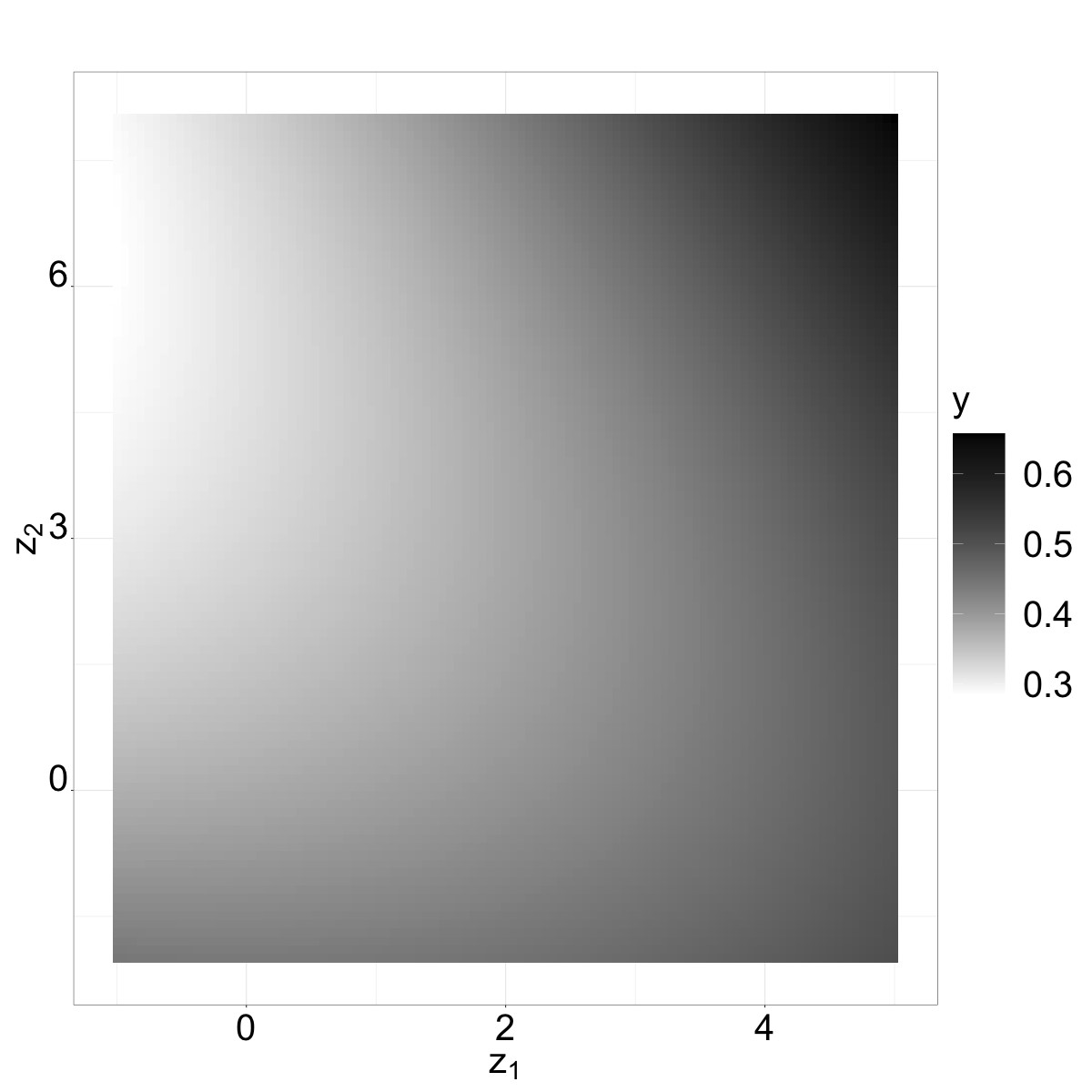}
		\caption{}
		\label{fig:real_1} 
	\end{subfigure}
	\hfill
	\begin{subfigure}[b]{0.475\textwidth}
		\includegraphics[width=1\linewidth]{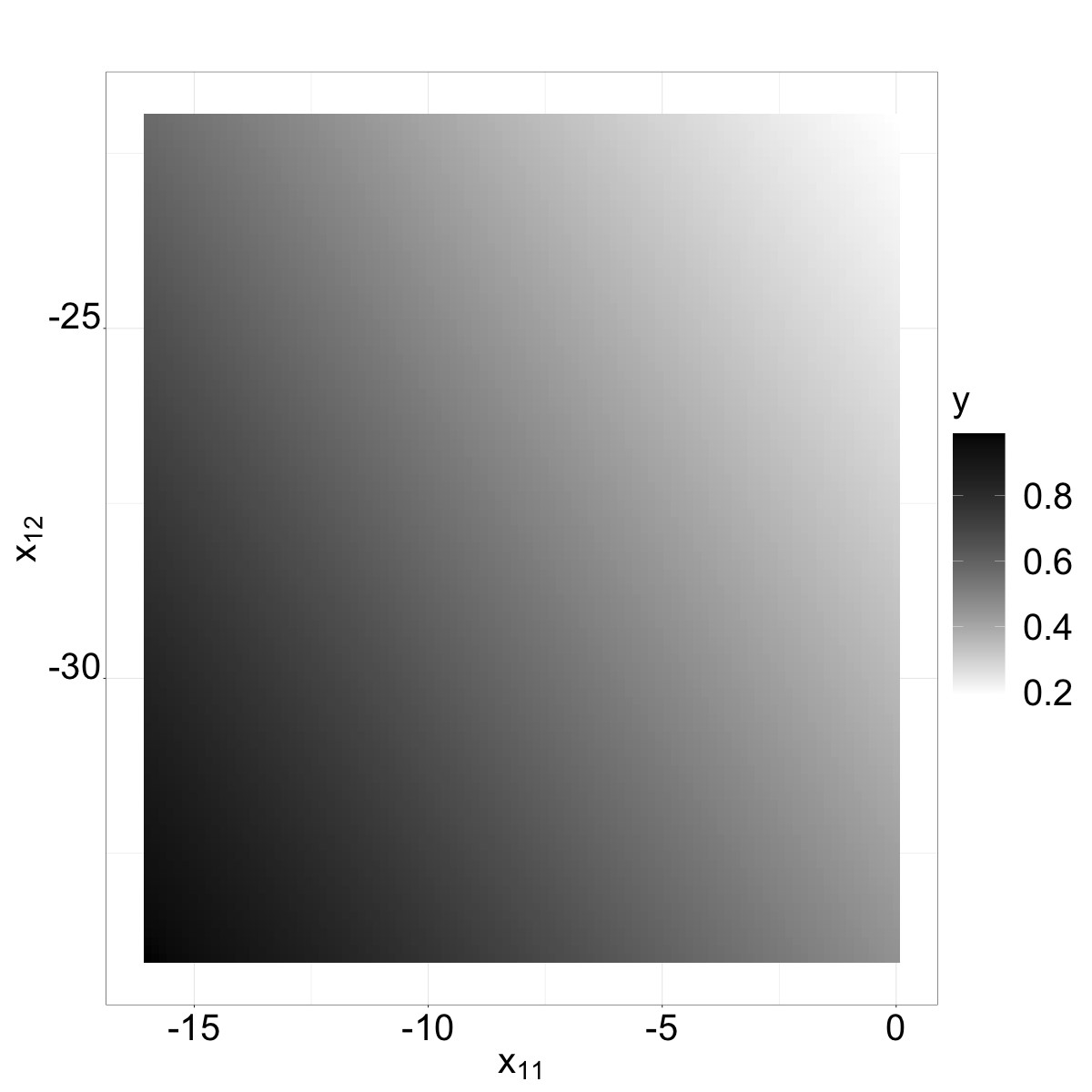}
		\caption{}
		\label{fig:real_2}
	\end{subfigure}
	\vskip\baselineskip
	\begin{subfigure}[b]{0.475\textwidth}
		\includegraphics[width=1\linewidth]{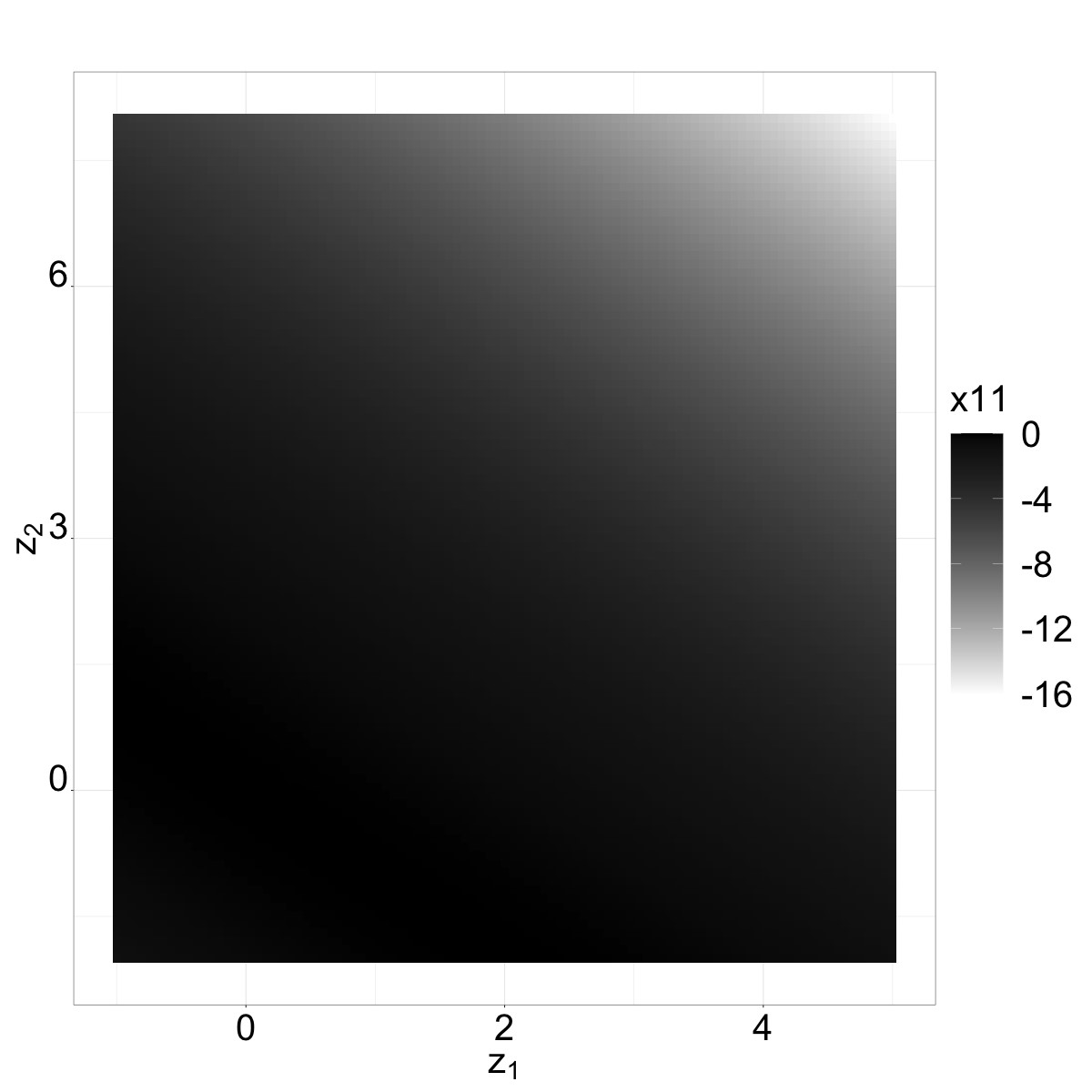}
		\caption{}
		\label{fig:real_3}
	\end{subfigure}
	\hfill
	\begin{subfigure}[b]{0.475\textwidth}
		\includegraphics[width=1\linewidth]{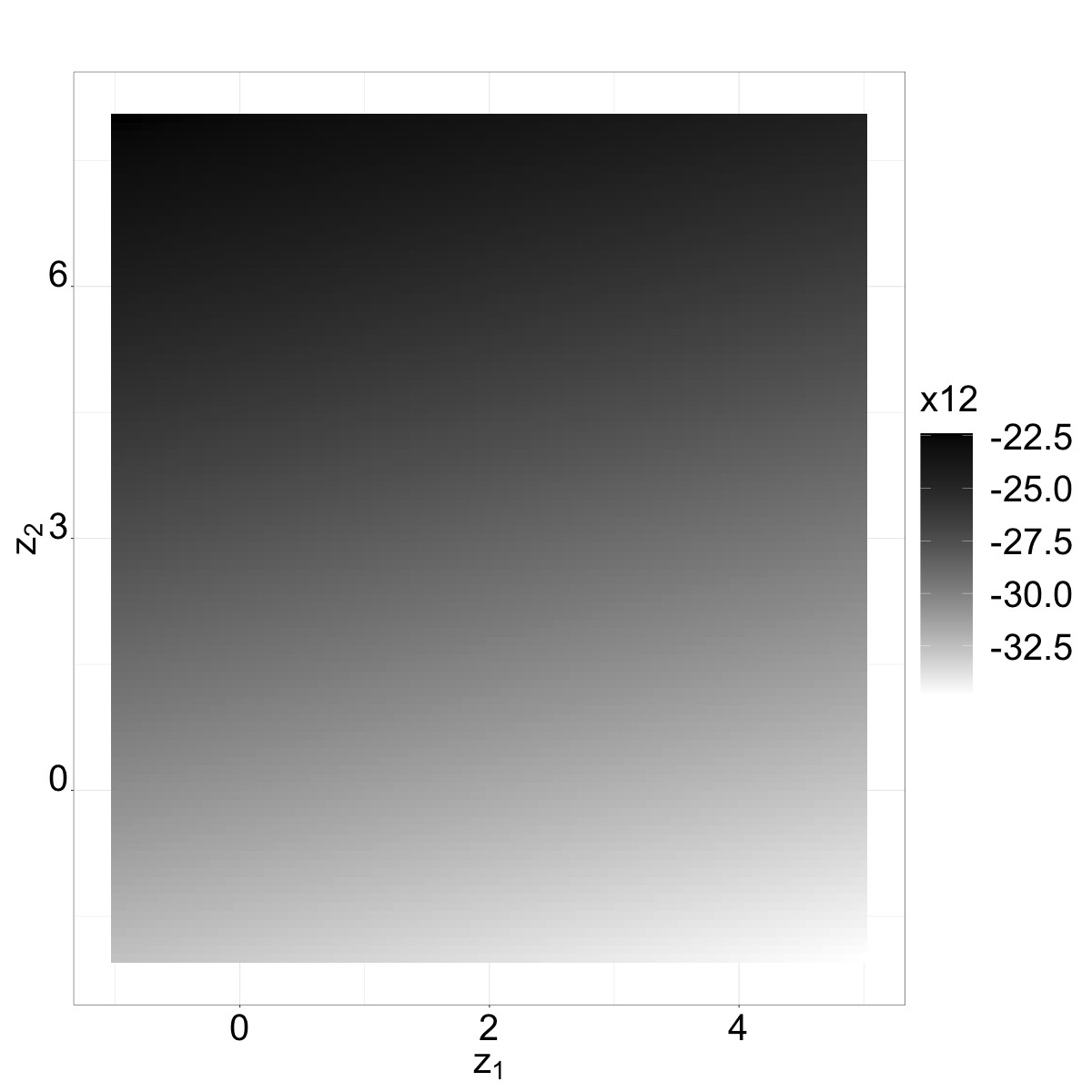}
		\caption{}
		\label{fig:real_4}
	\end{subfigure}
	
	\caption[TBD]{Heatmaps of the final selected model based on different perspectives to decompose its functional structure. In (a), the two-layer function $F(\boldsymbol{x}) = f^{(1)}_1\big[ f^{(1)}_{2}(\boldsymbol{x}), f^{(3)}_{2}(\boldsymbol{x}) \big]$ with input $\boldsymbol{x} = (z_{1}, z_2)$ is used to predict the outcome $y$. In (b), the output $y$ is estimated by the first layer function $f^{(1)}_1(\boldsymbol{x}_1)$ with input $\boldsymbol{x}_1 = (x_{1, 1}, x_{1, 2})$. In (c), $x_{1, 1}$ in $\boldsymbol{x}_1$ is estimated by a second layer function $f^{(1)}_{2}(\boldsymbol{x})$ with the original input $\boldsymbol{x}$. (d) is for $x_{1, 2}$ with $f^{(3)}_{2}(\boldsymbol{x})$. }
	\label{fig:real}
\end{figure}

Figure \ref{fig:real} provides a set of heatmaps to visualize the functional structure of the final selected model from different perspectives. In Figure \ref{fig:real_1}, the two-layer function $F(\boldsymbol{x}) = f^{(1)}_1\big\{ f_{2}^{(1)}(\boldsymbol{x}), f_{2}^{(3)}(\boldsymbol{x}) \big\}$ with input $\boldsymbol{x} = (z_{1}, z_2)$ is used to predict the outcome $y$. One can further decompose this relatively complex two-layer function by the first layer function $f_1^{(1)}(\boldsymbol{x}_1)$ with input $\boldsymbol{x}_1 = (x_{1, 1}, x_{1, 2})$ in Figure \ref{fig:real_2}, a second layer function $f_{2}^{(1)}(\boldsymbol{x})$ to estimate $x_{1, 1}$ with the original input $\boldsymbol{x}$ in Figure \ref{fig:real_3}, and $f_{2}^{(3)}(\boldsymbol{x})$ to estimate $x_{1, 2}$ with $\boldsymbol{x}$ in Figure \ref{fig:real_4}. The Figure \ref{fig:real_1} provides a neat single figure to illustrate the relationship between $y$ and $\boldsymbol{x}$, while Figures \ref{fig:real_2}, \ref{fig:real_3} and \ref{fig:real_4} offer sequential interpretation of impacts of $\boldsymbol{x}$ on intermediate variables $x_{1, 1}$, $x_{1, 2}$, and then the final outcome $y$, with base functions. 

In this section, we consider a setting with 2 elements in $\boldsymbol{x}$. This is mainly to facilitate visualization of the impact of these 2 variables on the outcome via heatmaps in Figure \ref{fig:real}. In the final selected model, the specific choice of 2 covariates (i.e., creatinine and HDL cholesterol) from a total of 45 combinations with a total of 10 potential explanatory variables is based on observed data. This setting can be generalized to more than 2 covariates, as implemented in Section \ref{sec:sim}. 

\section{Discussion}
\label{sec:disc}

In this article, we propose a general, flexible and harmonious framework to construct interpretable functions in regression analysis. Our method not only accommodates specific needs of interpretation, but also balances estimation, generalizability and simplicity. The resulting interpretable model facilitates downstream communications, motivates further investigations and increases transparency. 

There may exist identifiability issues of $\boldsymbol{\theta}$ in $F(\boldsymbol{x}; \boldsymbol{\theta})$ given all those candidate functions $\mathbb{F}_0$. This is an active and ongoing research topic on multi-layer functional models, such as DNN \citep{goodfellow2016deep}. In terms of interpretability, we are more interested in finding a feasible model with certain properties (e.g., interpretable functional form, well generalization ability), but not necessarily a unique one. Within a set of $\mathbb{F}_1$ or $\mathbb{F}_{2, j}$, $j = 1, ... J$, for a specific layer, it is relatively easier to choose identifiable base functions. A related question is how to characterize the uncertainty of the proposed framework. It generally involves two folds: the probability of selecting a specific model from candidates, and the randomness of parameter estimation in the selected model. One potential approach is to obtain empirical estimates of the above quantities based on Bootstrap samples. However, due to the relatively intensive computation of the proposed framework with DNN training, one can conduct future work to investigate a suggested Bootstrap sample size and potential computationally efficient methods. Another way to characterize uncertainty is to use the Bayesian framework, for example, a Bayesian DNN to quantify the uncertainty in DNN predictions \citep{zhang2023density}. 

This proposed framework has several future works. First of all, the scaler outcome $y$ can be extended to a vector, or more generalized outcomes, such as time series data or images. The objective function in (\ref{equ:f_k}) is able to incorporate additional penalized terms for variable selection. Additionally, the flexible model complexity parameter in modified Mallows's $C_p$-statistic in (\ref{equ:mcp}) can be generalized to other measures, such as expressive capacity \citep{hu2021model}. 

\section*{Acknowledgements}
The authors thank the Editor, the Associate Editor and two reviewers for their insightful comments to improve this article significantly.

This manuscript was supported by AbbVie Inc. AbbVie participated in the review and approval of the content. Tianyu Zhan is employed by AbbVie Inc., and may own AbbVie stock. Jian Kang's research is supported by NIH R01DA048993, NIH R01MH105561 and NSF IIS2123777. 

\section*{Supplementary Materials}

The R code to replicate results in Section \ref{sec:sim} and \ref{sec:real} is available on Oxford Academic and also on GitHub: \url{https://github.com/tian-yu-zhan/DNN_Interpretation}. 

\section*{Data Availability}

NHANES (National Health and Nutrition Examination Survey) data that support the findings in this paper and mentioned in Section \ref{sec:real} are available on \url{https://wwwn.cdc.gov/nchs/nhanes/search/datapage.aspx?Component=Laboratory&CycleBeginYear=2017}. 



%
\bibliographystyle{biom} 
\bibliography{In_ref.bib}

\begin{thebibliography}{}

\bibitem[\protect\citeauthoryear{Bauer and Kohne}{Bauer and
  Kohne}{1994}]{bauer1994evaluation}
Bauer, P. and Kohne, K. (1994).
\newblock Evaluation of experiments with adaptive interim analyses.
\newblock {\em Biometrics} pages 1029--1041.

\bibitem[\protect\citeauthoryear{Breiman, Friedman, Olshen, and J.}{Breiman
  et~al.}{1984}]{rt}
Breiman, L., Friedman, J.~H., Olshen, R.~A., and J., S.~C. (1984).
\newblock {\em Classification and Regression Trees}.
\newblock Wadsworthy.

\bibitem[\protect\citeauthoryear{Bretz, Koenig, Brannath, Glimm, and
  Posch}{Bretz et~al.}{2009}]{bretz2009adaptive}
Bretz, F., Koenig, F., Brannath, W., Glimm, E., and Posch, M. (2009).
\newblock Adaptive designs for confirmatory clinical trials.
\newblock {\em Statistics in medicine} {\bf 28,} 1181--1217.

\bibitem[\protect\citeauthoryear{Chollet and Allaire}{Chollet and
  Allaire}{2018}]{Chollet}
Chollet, F. and Allaire, J.~J. (2018).
\newblock {\em Deep Learning with R}.
\newblock Manning Publications Co., Greenwich, CT, USA.

\bibitem[\protect\citeauthoryear{Craven and Shavlik}{Craven and
  Shavlik}{1995}]{craven1995extracting}
Craven, M. and Shavlik, J. (1995).
\newblock Extracting tree-structured representations of trained networks.
\newblock {\em Advances in Neural Information Processing Systems} {\bf 8,}.

\bibitem[\protect\citeauthoryear{Doshi-Velez and Kim}{Doshi-Velez and
  Kim}{2017}]{doshi2017towards}
Doshi-Velez, F. and Kim, B. (2017).
\newblock Towards a rigorous science of interpretable machine learning.
\newblock {\em arXiv preprint arXiv:1702.08608} .

\bibitem[\protect\citeauthoryear{Floden and Bell}{Floden and
  Bell}{2019}]{floden2019imputation}
Floden, L. and Bell, M.~L. (2019).
\newblock Imputation strategies when a continuous outcome is to be dichotomized
  for responder analysis: a simulation study.
\newblock {\em BMC Medical Research Methodology} {\bf 19,} 1--11.

\bibitem[\protect\citeauthoryear{Gilmour}{Gilmour}{1996}]{gilmour1996interpretation}
Gilmour, S.~G. (1996).
\newblock {The interpretation of Mallows’s Cp-statistic}.
\newblock {\em Journal of the Royal Statistical Society Series D: The
  Statistician} {\bf 45,} 49--56.

\bibitem[\protect\citeauthoryear{Goodfellow, Bengio, and Courville}{Goodfellow
  et~al.}{2016}]{goodfellow2016deep}
Goodfellow, I., Bengio, Y., and Courville, A. (2016).
\newblock {\em Deep learning}.
\newblock MIT press.

\bibitem[\protect\citeauthoryear{Goodman and Flaxman}{Goodman and
  Flaxman}{2017}]{goodman2017european}
Goodman, B. and Flaxman, S. (2017).
\newblock European union regulations on algorithmic decision-making and a
  “right to explanation”.
\newblock {\em AI Magazine} {\bf 38,} 50--57.

\bibitem[\protect\citeauthoryear{Hastie, Tibshirani, Friedman, and
  Friedman}{Hastie et~al.}{2009}]{hastie2009elements}
Hastie, T., Tibshirani, R., Friedman, J.~H., and Friedman, J.~H. (2009).
\newblock {\em The elements of statistical learning: data mining, inference,
  and prediction}, volume~2.
\newblock Springer.

\bibitem[\protect\citeauthoryear{Hu, Chu, Pei, Liu, and Bian}{Hu
  et~al.}{2021}]{hu2021model}
Hu, X., Chu, L., Pei, J., Liu, W., and Bian, J. (2021).
\newblock Model complexity of deep learning: A survey.
\newblock {\em Knowledge and Information Systems} {\bf 63,} 2585--2619.

\bibitem[\protect\citeauthoryear{LeCun, Bengio, and Hinton}{LeCun
  et~al.}{2015}]{lecun2015deep}
LeCun, Y., Bengio, Y., and Hinton, G. (2015).
\newblock Deep learning.
\newblock {\em Nature} {\bf 521,} 436--444.

\bibitem[\protect\citeauthoryear{Li, Feng, Sui, Li, Song, Zhan, Cicconetti,
  Jin, Wang, Chan, et~al\mbox{.}}{Li et~al.}{2022}]{li2022analyzing}
Li, Y., Feng, D., Sui, Y., Li, H., Song, Y., Zhan, T., Cicconetti, G., Jin, M.,
  Wang, H., Chan, I., et~al. (2022).
\newblock Analyzing longitudinal binary data in clinical studies.
\newblock {\em Contemporary Clinical Trials} {\bf 115,} 106717.

\bibitem[\protect\citeauthoryear{Lipton}{Lipton}{2018}]{lipton2018mythos}
Lipton, Z.~C. (2018).
\newblock The mythos of model interpretability: In machine learning, the
  concept of interpretability is both important and slippery.
\newblock {\em Queue} {\bf 16,} 31--57.

\bibitem[\protect\citeauthoryear{Lundberg and Lee}{Lundberg and
  Lee}{2017}]{lundberg2017unified}
Lundberg, S.~M. and Lee, S.-I. (2017).
\newblock A unified approach to interpreting model predictions.
\newblock {\em Advances in Neural Information Processing Systems} {\bf 30,}.

\bibitem[\protect\citeauthoryear{Mallows}{Mallows}{1964}]{mallow1964}
Mallows, C.~L. (1964).
\newblock Choosing variables in a linear regression: A graphical aid.
\newblock {\em Central Regional Meeting of the Institute of Mathematical
  Statistics, Manhattan, KS} .

\bibitem[\protect\citeauthoryear{Nguyen, Yosinski, and Clune}{Nguyen
  et~al.}{2015}]{nguyen2015deep}
Nguyen, A., Yosinski, J., and Clune, J. (2015).
\newblock Deep neural networks are easily fooled: High confidence predictions
  for unrecognizable images.
\newblock {\em Proceedings of the IEEE conference on computer vision and
  pattern recognition} pages 427--436.

\bibitem[\protect\citeauthoryear{NHANES}{NHANES}{2024}]{NHANES}
NHANES (2024).
\newblock National health and nutrition examination survey.
\newblock \url{https://www.cdc.gov/nchs/nhanes/index.htm}.

\bibitem[\protect\citeauthoryear{Pearl}{Pearl}{2009}]{pearl2009causality}
Pearl, J. (2009).
\newblock {\em Causality}.
\newblock Cambridge university press.

\bibitem[\protect\citeauthoryear{Pulkstenis, Patra, and Zhang}{Pulkstenis
  et~al.}{2017}]{pulkstenis2017bayesian}
Pulkstenis, E., Patra, K., and Zhang, J. (2017).
\newblock A bayesian paradigm for decision-making in proof-of-concept trials.
\newblock {\em Journal of Biopharmaceutical Statistics} {\bf 27,} 442--456.

\bibitem[\protect\citeauthoryear{Rani, Liu, Sarkar, and Vanman}{Rani
  et~al.}{2006}]{rani2006empirical}
Rani, P., Liu, C., Sarkar, N., and Vanman, E. (2006).
\newblock An empirical study of machine learning techniques for affect
  recognition in human--robot interaction.
\newblock {\em Pattern Analysis and Applications} {\bf 9,} 58--69.

\bibitem[\protect\citeauthoryear{Ribeiro, Singh, and Guestrin}{Ribeiro
  et~al.}{2016}]{ribeiro2016should}
Ribeiro, M.~T., Singh, S., and Guestrin, C. (2016).
\newblock {"Why should iI trust you?" Explaining the predictions of any
  classifier}.
\newblock {\em Proceedings of the 22nd ACM SIGKDD International Conference on
  Knowledge Discovery and Data Mining} pages 1135--1144.

\bibitem[\protect\citeauthoryear{Ridgeway, Madigan, Richardson, and
  O'Kane}{Ridgeway et~al.}{1998}]{ridgeway1998interpretable}
Ridgeway, G., Madigan, D., Richardson, T., and O'Kane, J. (1998).
\newblock Interpretable boosted na{\"\i}ve bayes classification.
\newblock {\em Proceedings of the 4th International Conference on Knowledge
  Discovery and Data Mining} pages 101--104.

\bibitem[\protect\citeauthoryear{Wu, Hughes, Parbhoo, Zazzi, Roth, and
  Doshi-Velez}{Wu et~al.}{2018}]{wu2018beyond}
Wu, M., Hughes, M., Parbhoo, S., Zazzi, M., Roth, V., and Doshi-Velez, F.
  (2018).
\newblock Beyond sparsity: Tree regularization of deep models for
  interpretability.
\newblock {\em Proceedings of the AAAI Conference on Artificial Intelligence}
  {\bf 32,}.

\bibitem[\protect\citeauthoryear{Zhan}{Zhan}{2022}]{zhan2022dl}
Zhan, T. (2022).
\newblock {DL 101: Basic introduction to deep learning with its application in
  biomedical related fields}.
\newblock {\em Statistics in Medicine} {\bf 41,} 5365--5378.

\bibitem[\protect\citeauthoryear{Zhan and Kang}{Zhan and
  Kang}{2022}]{zhan2022finite}
Zhan, T. and Kang, J. (2022).
\newblock Finite-sample two-group composite hypothesis testing via machine
  learning.
\newblock {\em Journal of Computational and Graphical Statistics} {\bf 31,}
  856--865.

\bibitem[\protect\citeauthoryear{Zhang, Liu, and Kang}{Zhang
  et~al.}{2023}]{zhang2023density}
Zhang, D., Liu, T., and Kang, J. (2023).
\newblock Density regression and uncertainty quantification with bayesian deep
  noise neural networks.
\newblock {\em Stat} {\bf 12,} e604.

\bibitem[\protect\citeauthoryear{Zhang, Ti{\v{n}}o, Leonardis, and Tang}{Zhang
  et~al.}{2021}]{zhang2021survey}
Zhang, Y., Ti{\v{n}}o, P., Leonardis, A., and Tang, K. (2021).
\newblock A survey on neural network interpretability.
\newblock {\em IEEE Transactions on Emerging Topics in Computational
  Intelligence} {\bf 5,} 726--742.

\end{thebibliography}

\end{document}